\documentclass[usenatbib]{article}
\usepackage{times,graphicx,natbib,a4wide}

\begin{document}

\title{The Entropy Principle, and the Influence of Sociological Pressures on SETI}
\author{V. Bozhilov$^1$ and Duncan H. Forgan$^2$}
\maketitle

\noindent $^1$Faculty of Physics, Department of Astronomy, Sofia University, Bulgaria; email: archivl@yahoo.com \\
\noindent $^2$Scottish Universities Physics Alliance (SUPA), Institute for Astronomy, University of Edinburgh; email: dhf@roe.ac.uk 

\begin{abstract}

\noindent We begin with the premise that the law of entropy could prove to be fundamental for the evolution of intelligent life and the advent of technological civilization. Building on recent theoretical results, we combine a modern approach to evolutionary theory with Monte Carlo Realization Techniques. A numerical test for a proposed significance of the law of entropy within the evolution of intelligent species is performed and results are compared with a neutral “test” hypothesis. Some clarifying aspects on the emergence of intelligent species arise and are discussed in the framework of contemporary astrobiology. \\
 

\end{abstract} 

\newpage

\section{Introduction}

\noindent Although panspermia \citep{Hoyle1977} and the Rare Earth hypothesis \citep{rare_Earth} are two of the most popular paradigms regarding the evolution of ETIs, modern biology indicates that alternative theories on the development of intelligent life may also carry weight. Recent papers \citep{JaakkolaSalla2008,JaakkolaSalla2009,Kaila2008,Fu2007,NamBozhilov2009} have demonstrated that up until now, one fundamental principle of nature – the second law of thermodynamics, has not yet been fully incorporated in evolutionary algorithms. \\

\noindent Entropy is fundamental to thermodynamics and its importance in physics and in astrophysics is indisputable. Biology is now incorporating it into the evolutionary paradigm. Numerous analyses (ibid.) show evolution might be understood as a process of constant complexity increase, further developing the correlation between entropy and evolution on both microscopic and macroscopic scales. \\

\noindent Therefore, it may prove worthwhile to study the effect of the 2nd Law on the latter stages of evolution – in particular the evolution of intelligent technological civilizations. This paper aims to explore a new hypothesis regarding the emergence of intelligent life, which views the process of natural selection and evolution as intrinsically linked with the second law of thermodynamics.  \\

\noindent The paper is divided into 6 sections. The “Entropy Hypothesis”, which links the second law of thermodynamics with the evolution of intelligent life, is postulated in section 2.  In section 3, we outline the numerical method by which the development of life and intelligence in the Galaxy can be simulated.  In section 4, we display the results of numerical simulations describing the development of “entropy-driven” ETIs in the Galaxy. This data is compared with a neutral “baseline” simulation to study the unique features of “entropy-driven” ETIs.  We discuss the results and provide conclusions in sections 5 and 6 respectively.  \\

\section{The Evolution of Intelligent Technological Civilizations, the Second Law of Thermodynamics and the Entropy Hypothesis}

\noindent Life on Earth began approximately some 3.8 Gyr ago \citep{Mojzsis1996,Ehrenfreund2002,Manning2006}. The evolution towards intelligent life comprises 6 important stages \citep{stages}: biogenesis, the advent of bacteria, the advent of eukaryotes, combigenesis, the advent of metazoans, and the birth of technological civilization.  \\

\noindent Evolution is “boosted” as these stages are achieved, giving rise to a “step-like” scenario. Some of these steps are “hard” in the evolutionary sense i.e. – the probability of their occurrence is low, and therefore require relatively long timescales to occur. The biogenesis and the emergence of eukaryotes are thought as the two “hard steps” in the evolutionary scenario \citep{stages}. However, the many possible evolutionary pathways that lead to the development of intelligent life forms and a technically advanced civilization like ours, remain difficult to define (\citealt{Cirkovic2007,Spiegel_et_al_08,Lal2008} and references within).  \\

\noindent According to recent theory, \textbf{the Second Law of Thermodynamics could prove to be essential for understanding biological evolution}.  The mathematical analysis of the importance of the Second Law in evolution is stated in \citet{Kaila2008}: \\

\noindent \emph{“The second law of thermodynamics is a powerful imperative that has acquired several expressions during the past centuries. Connections between two of its most prominent forms, i.e. the evolutionary principle by natural selection and the principle of least action, are examined. Although no fundamentally new findings are provided, it is illuminating to see how the two principles rationalizing natural motions reconcile to one law.”} \\

\noindent The equation of evolution including entropy can be used \citet{JaakkolaSalla2008} to explain the differences in the genome, as a consequence of the second law of thermodynamics. Others take the problem to a macroscopic scale and demonstrate that the same technique can explain the species-area relationship, one of eco-biology’s key problems \citet{WurtzPeter2009}.  Indeed, it is possible that the universal criterion for evolutionary selection is the entropy principle \citep{Sharma2007123,JaakkolaSalla2009}.  Taking into account the reach of the law of entropy over various scales, we now summarize what we name “The Entropy Hypothesis”:   \\

\noindent \emph{The intelligent technological civilizations are a typical (although not guaranteed) consequence of the biological evolution of complex life forms, provided the necessary conditions are met.  This is due to the efficiency of technological civilizations at increasing the entropy of their planetary system on very short timescales, satisfying the 2nd Law of Thermodynamics. The destruction of such a technological civilization, which may be inherent in their evolution, will in general be the most effective way for biological evolution to fulfill the law of Entropy. So, whenever the conditions for evolution of complex life forms towards intelligence are met, an intelligent technological civilization will appear, constantly evolve technically until it is self-destructed, colonized by another civilization, or starts colonizing space itself, thus ensuring the increase of entropy on even larger scales.} \\

\noindent What is the basis for this hypothesis? Life and living organisms in a closed ecosystem decrease entropy. However, the organisms that survive \citep{Sharma2007123, JaakkolaSalla2009, Fu2007} are those that absorb the free energy as effectively as possible. Still, the entropy can be increased globally, if there is a way to alter ecosystems on large size scales and short timescales.   \\

\noindent For the approx. $10^5$ years since the emergence of homo sapiens, Mankind has developed technology that can affect the Earth globally (in particular through the construction of buildings and deforestation destroying habitats). Much of Earth’s surface has been altered in a  short amount of (cosmic) time. If our technology continues to evolve and/or a technological breakdown occurs, vast amounts of the planet could easily be destroyed or contaminated. From technology, therefore, a sociological pressure is derived: intelligent life is capable of self-destruction.  This is true for Mankind, and we assume it to be true for other technological civilizations; \textbf{we argue this is a general effect of the second law of thermodynamics on macro-scale}.  \\

\noindent As an efficient entropy generator, we could suppose technology is connected to and would evolve shortly after the development of an intelligent species on a given planet, giving a “natural” mechanism for rapidly increasing the entropy on a planetary scale, as an extension of the law of entropy guiding natural selection. This would suggest that the intelligent technological civilizations could no longer be thought of as an exceptional or even rare “event” in biological evolution.  That is not the only avenue which evolution can take, but its effect on increasing the entropy (we argue) may favor it over other potential avenues. \textbf{Thus, we have theoretical expectations on the routes evolution can take on other biospheres we might find in space, other than Earth. On planets that are Earthlike, we expect that while the initial biochemistry of life may be very different from Earth’s, the selection pressures introduced by the environment will be similar, possibly resulting in convergent evolution (e.g. \citealt{Morris2006R826}).}   \\

\noindent \textbf{Furthermore, we can try to elaborate a definition of intelligence in the framework of the second law of thermodynamics. Human evolution might be regarded as survival of the fittest replicators, where by replicators we denote organisms, devices or even concepts (e.g. memes, \citealt{Dawkins1990}) that can reproduce themselves and are subject in some form to natural selection.  The first (biochemical) replicators are the genes, or more specifically RNA and DNA. As has been argued by previous authors (ibid.) that these replicators that survive during natural selection are namely the ones that absorb free energy most effectively, i.e. natural selection is directly interrelated with the entropy principle. We can identify technology and culture as crucial milestones in the development of Man as a sentient species.  These can be thought of as replicators that have been artificially synthesized by humans, or (to take the neo-Darwinian view) the genes themselves.  Thus, given the entropy hypothesis of biological evolution, we can make a tentative definition of intelligence:} \\

\noindent \emph{Intelligence is the process by which replicators artificially synthesize a radically new and fundamentally different type of replicator.} \\

\noindent \textbf{As replicators are subject to natural selection (and therefore the entropy principle by extension), this definition encourages us to regard intelligence as a standard effect in evolution, which arises in order to assure the 2nd law of thermodynamics on a macro scale.  This definition can also be used in other disciplines such as sociology, economics and biology, allowing us to gain new insights and deepen our understanding of intelligence as a natural paradigm (see also \citealt{Kaila2008}).}  \\

\noindent \textbf{Intelligence defined in this way is not restricted to biological replicators alone.  Other replicators also have the potential to satisfy our criterion for intelligence, such as machines, provided these replicators synthesize other ones without guided supervision (e.g. without intervention of external intelligent observers).  Indeed, it suggests that for machines to become truly intelligent, they must be sufficiently developed to become true replicators subject to the laws of natural selection. } \\

\noindent Once an intelligent civilization has arisen, it can either self-destruct, or alternatively, provided it survives long enough and develops the necessary technology, it can begin colonizing other nearby planets, thus still increasing entropy at a maximum possible rate, by changing and/or destroying another worlds. However, to simplify this analysis, colonization effects will not be considered in this work.  \\

\noindent In brief, in the frame of current results, it seems plausible that the law of entropy could be a primary cause for the development of intelligent species and a key factor for the advent of technological civilization, regarded as a natural mechanism, assuring the quickest possible entropy rate increase in a given planetary ecosystem. 

\section{Numerical Method and Parameters}

\noindent We do not have any current observations for the presence of another technological civilization. Hence, we must turn to numerical methods - in this case, Monte Carlo Realization techniques (see also \citealt{Vukotic_and_Cirkovic_07, Vukotic_and_Cirkovic_07}).  These can be used to statistically model our Galaxy and, as a first approximation, estimate the possible number of ETIs in the Milky Way. This is done in \citet{mcseti}.  In present work, we apply some changes to the model (described in \citealt{mcseti2}) to test the entropy hypothesis alongside a neutral baseline hypothesis.  To aid the reader, we briefly describe those methods here, \textbf{but suggest reading the articles cited for a detailed description.}  \\

\noindent In essence, the method generates a Galaxy of N* stars, each with their own stellar properties (mass, luminosity, location in the Galaxy, etc.). Planetary systems are then generated for these stars, and life is allowed to evolve on these planets according to some hypothesis of origin.   The life on these worlds is evolved individually according to stochastic equations of evolution using the “hard step paradigm”, \textbf{described in section 2 (see \citealt{mcseti} for details)}.  Intelligent life can then form on these worlds if they can survive resetting events such as asteroid impacts or gamma ray bursts (e.g. \citealt{Annis}).  The end result is a mock Galaxy which is (astrophysically speaking) statistically representative of the Milky Way, with the addition of life and intelligent species. To quantify random sampling errors, this process is repeated many times. This allows an estimation of the sample mean and sample standard deviation of the output variables obtained.  \\

\noindent The inputs used to define the mock Galaxy (e.g. the Galaxy’s surface density profile, the initial stellar mass function (IMF), the star formation history (SFH), etc.) are of critical importance.  \citet{mcseti} focused on using current empirical data (especially for the simulation of exoplanets) to define the mock Galaxy.  \citet{mcseti2} outline improvements to the model which allows the code to correctly simulate Earth-like planets.  
The star formation history and age metallicity relation used in this model can be found in \citep{Rocha_Pinto_AMR,Rocha_Pinto_SFH}; the initial mass function is taken from \citet{Miller1979}.  The planets are selected from the same distribution functions as used in \citet{mcseti2}. The reader is referred to that work (as well as \citealt{mcseti}) for more information and numerical details.  \\

\noindent Two separate hypotheses were tested with this model, which are described below.

\subsection{The Baseline Hypothesis}

\noindent This neutral hypothesis is required for effective comparison, as the numerical method is better at determining relative trends than absolute values \citep{mcseti,mcseti2}. The baseline hypothesis requires only that a planet is in the stellar habitable zone for life to form upon it. If the planet’s surface temperature lies between [0, 100]◦C, then microbial life can form. Complex animal life will only form if the planet’s surface temperature lies between [4, 50]◦C \citep{rare_Earth}.

\subsection{The Entropy Hypothesis}

\noindent Finally, we introduce the effect predicted by the entropy hypothesis. We rewrite the 	parameter describing the self-destruction probability as:
￼
\begin{equation} P_{destroy} = 1.65 \times 10^{-3}e^{\frac{t_{adv}}{0.056}} \end{equation}

\noindent This is done so that Pdestroy ranges from 0.01 to 0.9 across all possible $t_{adv}$ values (where $t_{adv}$ is the timescale for a civilization to move beyond its fledgling stage, and escape self-destruction). $t_{adv}$ is selected from a Gaussian with mean 0.25 Gyr, and standard deviation of 0.1 Gyr. This approach represents the sociological pressure, marking the possibility of destruction (the “preferred” maximum entropy state) becoming more probable with technological advance.  By comparison, the baseline sets $P_{destroy} = 0.5$ for all civilizations, reflecting “ignorance” as to the sociological issues of each individual civilization.

\section{Results}

\noindent Each hypothesis was run for 30 separate Monte Carlo Realizations (MCRs) requiring 144 CPU hours each.  Separate analysis of the connectivity of civilization pairs took 1152 hours for each hypothesis. A total of around 2600 CPU hours were required to produce this data\footnote{Note: Errors are plotted on all graphs in this paper.  Even though the errors seem small, the results of the two hypotheses should be considered relative to each other. The true error in the parameterization of the Galaxy is not incorporated in this calculation \citep{mcseti,mcseti2}}.

 

\noindent For each planet in the simulation, a habitation index is given, according to its evolutionary history. The meaning of each index is as follows:
￼
\begin{equation} I_{inhabit} = \left\{
\begin{array}{l l }
-1 & \quad \mbox{Biosphere which has been annihilated} \\
0 & \quad \mbox{Planet is lifeless} \\
0.5 & \quad \mbox{Planet has microbial life} \\
1 & \quad \mbox{Planet has primitive animal life} \\
2 & \quad \mbox{Planet has intelligent life} \\
3 & \quad \mbox{Planet had intelligent life, but it destroyed itself} \\
4 & \quad \mbox{Planet has an advanced civilisation} \\
\end{array} \right. \end{equation}

￼\begin{figure}
\begin{center}
\includegraphics[scale=0.7]{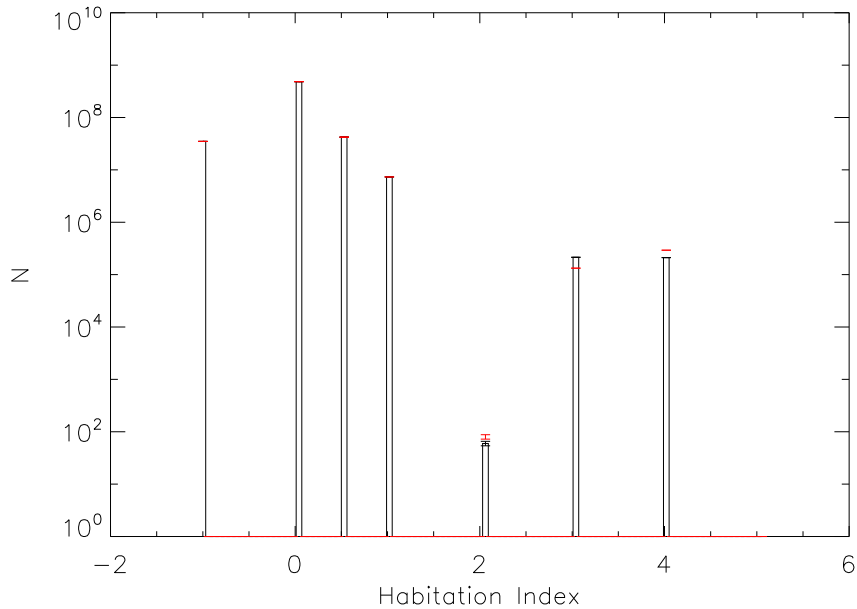}
\caption{Habitation index for the Entropy (red) and The Baseline (black) Hypothesis.  The habitation index legend is given in the text.\label{fig:inhabit}}
\end{center}
\end{figure}

\noindent The breakdown of habitation index for both hypotheses is shown together in Figure \ref{fig:inhabit}. No significant change is observed: most of the planets in the simulation never develop intelligent life ($-1\leq I\leq< 1$). The planets with index $I_{inhabit}=2$ i.e. the planets with “young” technological civilizations or “fledgling” civilizations, are relatively the same number (around 102). There is a small change in the number of fledgling civilizations which destroy them-selves ($I_{inhabit}=3$). The Entropy hypothesis produces a smaller number of self-destructed young ETIs, which gives rise to a higher number of advanced ($I_{inhabit}=4$) technological civilizations instead. As expected, the Entropy hypothesis as characterized in section 3.2 affects only the latter stages of evolution and technological development (which seems to be more prevalent against the baseline).  This is in concordance with the expectation that the ETI’s evolution is entropy-driven, which results in a “innovate or die” sociological pressure.  \\

\noindent The Entropy Hypothesis does not speculate on the location of Life in the Galaxy, so it is expected to match the results of the baseline. This is indeed the case: there is no change in the galactocentric radius of the planets for both hypotheses, as can be seen in Figure 2.  Both hypotheses reproduce as expected the Galactic Habitable Zone \citep{GHZ} at around 8 kpc, demonstrating the balance between Galactic chemical gradients and potentially sterilizing astrophysical phenomena. \\

\begin{figure}
\begin{center}
\includegraphics[scale=0.7]{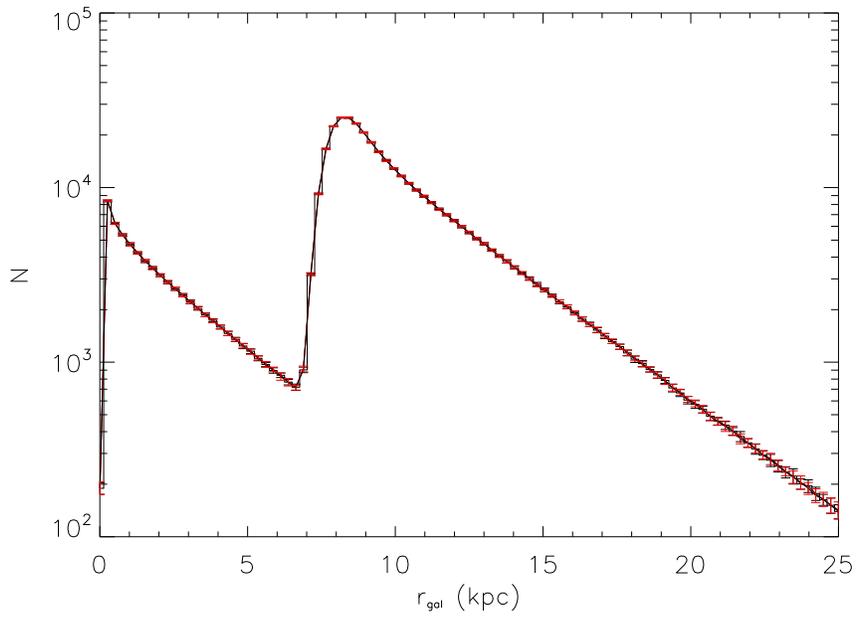}
\caption{The number of planets (N) and their galactocentric radius rgal in kpc for the Entropy (red) and Baseline (black) hypothesis.  \label{fig:rgal}}
\end{center}
\end{figure}
￼
\noindent \textbf{We can define the signal history of the Galaxy as the total number of communicating civilizations as a function of time.} Will this history identify the influence of entropy’s sociological pressure on ETIs?  Figure 3 shows that the Entropy hypothesis gives a slight enhancement against the baseline; the increase in advanced civilizations means their signal lifetime is longer, enhancing their number N over a more significant period of cosmic time.  However, this enhancement is within the error bars of the baseline measurement. This would suggest that confirming the entropy hypothesis by observing N alone would be hazardous if not impossible (not least because N is currently measured as 1).  \\

\begin{figure}
\begin{center}
\includegraphics[scale=0.7]{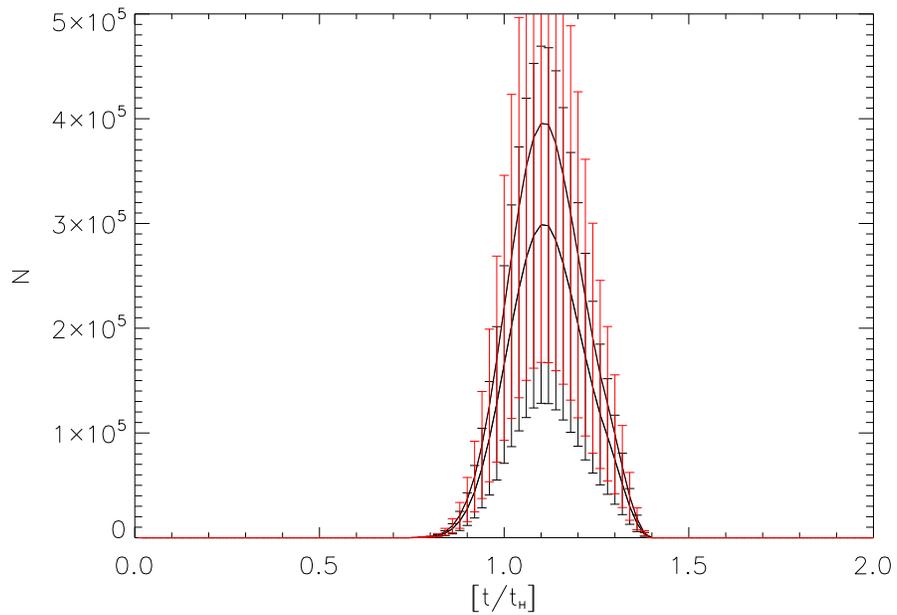}
\caption{The signal history of the Milky way, normalized by the Hubble time $t_H$. $t=t_H$ represents the present day. The baseline hypothesis is shown in black and the entropy hypothesis in red, along with the proper error bars.  \label{fig:signal}}
\end{center}
\end{figure}

\noindent If the increased signal lifetime is responsible for this slightly enhanced signal history, then this should be quantified.  The lifetime of the emitted signals for each hypothesis is shown in Figure 4. Although there is a significant increase in the number of signals, this occurs only to signals with lifetime around 1-3 Gyr. No significant change is observed for longer times. \\
￼
\begin{figure}
\begin{center}
\includegraphics[scale=0.7]{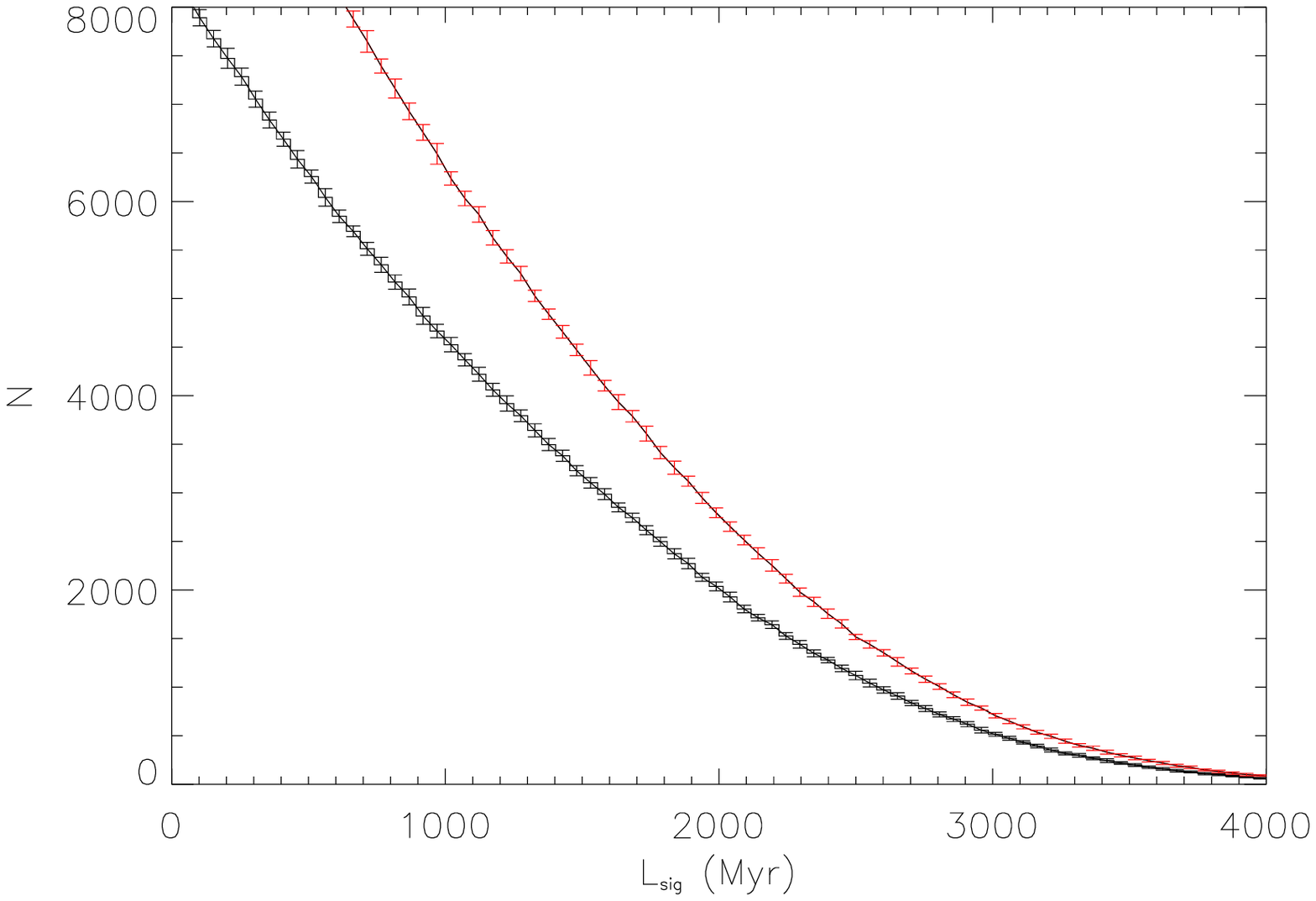}
\caption{The lifetime of emitted (leaked) ETI signals, relative to the number of signals.  The Baseline hypothesis is shown in black and the Entropy hypothesis in red. The count of the longest long-living signals (over than 4 Gyr), which are important for communication over very long distances (naturally of most interest to SETI researchers) remain unchanged. A considerable increase in number occurs in the signals with shorter lifetimes (1-2 Gyr).  \label{fig:Lsig}}
\end{center}
\end{figure}

\noindent It is difficult to place any constraints on SETI based on studies of this nature: analyzing the connectedness of these civilizations is the most concrete means at our disposal.  Figure 5 displays the results of the “contact factor” (the number of conversations or pairs of signals that any pair of civilizations can exchange, \citealt{mcseti2}).  Most of the simulated planets, even if they host ETI, are disconnected: either the distance is too great, or the time interval in which the pair co-exists is too short.   The Entropy hypothesis tends to produce civilizations that are more connected, with a higher contact factor (this is to be expected if the Entropy hypothesis produces more advanced civilizations on average than the baseline).   \\
  
\begin{figure}
\begin{center}
\includegraphics[scale=0.7]{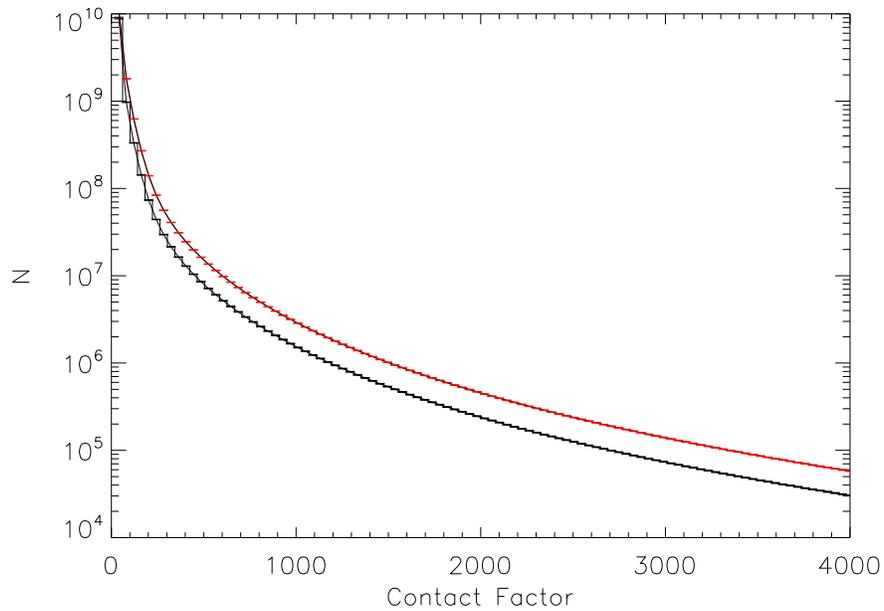}
\caption{ Results of the contact factor (measuring the number of successfully exchanged signals) are plotted. The baseline hypothesis is shown in black, the entropy hypothesis in red.  For most of the signals, no change observed. The significant increase in successfully transmitted communication occurs only to a small number of pairs at high contact factors \label{fig:c_fac}}
\end{center}
\end{figure}
  
\noindent \textbf{Note that the contact factor is dependent on the space separation between each inhabited planet and the time interval for each civilization pair. The entropy hypothesis yields changes only to the time interval, but not to the physical separation, i.e. the distance between the host planets of alien civilizations. Also, our estimates of connectivity make no assumptions about the methods by which communication is established.  Traditionally SETI has favoured radio  emission, but this particular method may restrict connectivity much more strongly than was previously realized (Forgan and Nichol, in prep.).  It is plausible that civilizations (including ours) will employ other communication techniques, e.g. EM radiation at other wavelengths, or more exotic communications methods based on less attenuating phenomena like neutrinos \citep{LearnedJ.G.1994,SilagadzeZ.K.2008}. However, increasing connectivity by this method will require more free energy, perhaps discouraging civilizations to communicate when resources become scarce \citep{empire}.} \\

\section{Discussion}

\noindent The entropy principle is connected with the actions and the sociological behavior of ETI, so it becomes decisive only at a later evolutionary stage e.g. after technology is developed. Note that in this analysis we implicitly suppose that intelligent species always discover technology if they survive long enough. \textbf{The actual behavior of technological civilizations might differ from our expectations. } \\

\noindent \textbf{In fact, intelligent civilizations may behave more like viruses, expanding and optimising their free energy consumption to attain a steady state if there are sufficient populations of hosts (i.e. biospheres) available (Starling \& Forgan, in prep.).  There is a well-defined period in which the technological civilizations are highly susceptible to destruction (the ``fledgling period" in our numerical simulations where civilizations progress from $I_{inhabit}=2$ to $I_{inhabit}=3$ or $I_{inhabit}=4$). Nevertheless, the tendency towards expansion (i.e. consuming free energy available) will ultimately lead the civilization first to harvest the energy of its planet, then the energy of the host star and finally - may be even the energy of entire galaxy. Thus, by the use of the entropy principle, we motivate theoretically the Kardashev scale for classifying technological civilizations based on their energy consumption \citep{KardashevN.S.1964}.} \\

\noindent The numerical results exhibit a clear tendency of favoring the evolution of increasingly advanced technology (see Figure 1). The Entropy hypothesis speculates this could be due to the inherited effect of the entropy principle: the ETIs “are bound”, if possible and under the right environmental conditions, to discover and evolve technology as a new source of increasing complexity, until either self-destruction or progression to a stable, advanced state occurs. Thus, we state that intelligent species may be regarded as the most effective way of securing the fast and constant increase of the rate of entropy, considering a given planetary ecosystem. \\

\noindent However, given the observational data available, there is no current means of proving or disproving the entropy hypothesis. The potential for future radio telescopes such as the Square Kilometer Array (SKA) to map out artificial signals is limited to a region less than 300 light years from the Earth \citep{Loeb2007}, and the combined decay of signal leakage from the Earth suggests that even the SKA will not be efficient at detecting ETI (Forgan \& Nichol, in prep).  Even in the distant future, when a more detailed Galactic census is available to us, we may not be able to distinguish the effects of entropy-driven evolution from a more neutral hypothesis.  \\

\noindent Still, if intelligence could be reasoned as a normal stage in evolution, as expected by the entropy hypothesis, there might well be a larger number of emitted or successfully exchanged signals between the different ETIs.  The results demonstrate that although there is indeed a minor increase in expected civilizations in the present day (Figure 3), it will be difficult to observe this increase (or the increased connectivity suggested in Figures 4 and 5).  While the entropy mechanism provides an appealing explanation of the emergence of intelligent species, current SETI-type observations will not provide sufficient evidence to be able to establish its veracity.  \\

\noindent Conversely, the entropy hypothesis provides an (often-cited) answer to the Fermi paradox - we do not see alien life, because ETIs tend to have a shorter lifetime than we might naively expect.  In addition, by incorporating the second law of thermodynamics into recent astrobiological analysis, we may be able to uncover more details on the origin of civilization, and its interaction with Earth’s ecosystem.

\section{Conclusions}

\noindent The entropy principle is fundamental in modern evolutionary theory: natural selection and biological evolution are deeply correlated and can be understood within the second law of thermodynamics. The direction of genomic evolution, just as other evolutionary processes, toward more probable distributions can be deduced from the logarithmic probability measure known as entropy \citep{Sharma2007123,Kaila2008,JaakkolaSalla2009,Fu2007}.   \\

\noindent We investigated the extension of this principle into intelligent technological civilizations through Monte Carlo Realization techniques. By comparing this entropy hypothesis with a more neutral baseline, it can be seen that the entropy hypothesis drives the creation and development of technology (as a new resource of increasing complexity).  This enhances the number of technological civilizations that reach an advanced stage of evolution (under a sociological pressure of “innovate or die”).  \\

\noindent  While the emergence of intelligent species might not be inevitable, but the process of creating such “replicators” in evolution is likely to be favoured, because it assures the maximum amount of free energy and will generate more entropy. \textbf{This leads to a general definition of intelligence based on entropy arguments (discussed in section 2), which provides further insight into evolutionary theory}. However, the detectability of this sociological pressure is low: in general, although the entropy hypothesis deviates slightly from the baseline data, this deviation is too small to be able to distinguish it and prove it observationally using current instruments.   \\

\noindent This work highlights two important points: firstly, that the influence of entropy may go further than was previously thought, especially in understanding the evolution of intelligent species. Secondly, our results demonstrate that sociological pressure will affect (if only slightly) future development of the Galactic community of civilizations over cosmic time, and should be considered in further studies of this type.

\section{Acknowledgements}

\noindent This work has made use of the resources provided by the Edinburgh Compute and Data Facility (ECDF, http://www.ecdf.ed.ac.uk/). The ECDF is partially supported by the eDIKT initiative (http://www.edikt.org.uk).

\bibliographystyle{mn2e} 
\bibliography{entropy}

\begin{thebibliography}{}

\bibitem[\protect\citeauthoryear{Annis}{Annis}{1999}]{Annis}
Annis J.,  1999, J. Br. Interplanet. Soc., 52, 19

\bibitem[\protect\citeauthoryear{Carter}{Carter}{2008}]{stages}
Carter B.,  2008, International Journal of Astrobiology, 7, 177

\bibitem[\protect\citeauthoryear{\'{C}irkovi\'{c}}{\'{C}irkovi\'{c}}{2007}]{Ci%
rkovic2007}
\'{C}irkovi\'{c} M.~M.,  2007, International Journal of Astrobiology, 6, 325

\bibitem[\protect\citeauthoryear{Cirkovic}{Cirkovic}{2008}]{empire}
Cirkovic M.~M.,  2008, Journal of the British Interplanetary Society, 61, 246

\bibitem[\protect\citeauthoryear{Dawkins}{Dawkins}{1990}]{Dawkins1990}
Dawkins R.,  1990, {The Selfish Gene}.
Oxford University Press, USA

\bibitem[\protect\citeauthoryear{Ehrenfreund, Irvine, Becker, Blank, Brucato,
  Colangeli, Derenne, Despois, Dutrey, Fraaije, Lazcano, Owen, Robert \&
  ISSI-Team}{Ehrenfreund et~al.}{2002}]{Ehrenfreund2002}
Ehrenfreund P.,  Irvine W.,  Becker L.,  Blank J.,  Brucato J.~R.,  Colangeli
  L.,  Derenne S.,  Despois D.,  Dutrey A.,  Fraaije H.,  Lazcano A.,  Owen T.,
   Robert F.,    ISSI-Team a. I. S. S.~I.,  2002, Reports on Progress in
  Physics, 65, 1427

\bibitem[\protect\citeauthoryear{Forgan}{Forgan}{2009}]{mcseti}
Forgan D.,  2009, International Journal of Astrobiology, 8, 121

\bibitem[\protect\citeauthoryear{Forgan \& Rice}{Forgan \&
  Rice}{2010}]{mcseti2}
Forgan D.,  Rice K.,  2010, International Journal of Astrobiology, 9, 73

\bibitem[\protect\citeauthoryear{Fu}{Fu}{2007}]{Fu2007}
Fu S.,  2007, Arxiv e-prints 0712.2108

\bibitem[\protect\citeauthoryear{Hoyle \& Wickramasinghe}{Hoyle \&
  Wickramasinghe}{1977}]{Hoyle1977}
Hoyle F.,  Wickramasinghe N.~C.,  1977, Nature, 268, 610

\bibitem[\protect\citeauthoryear{{Jaakkola, Salla}, {El-Showk, Sedeer} \&
  {Annila, Arto}}{{Jaakkola, Salla} et~al.}{2008}]{JaakkolaSalla2008}
{Jaakkola, Salla} {El-Showk, Sedeer}   {Annila, Arto} 2008, eprint
  arXiv:0807.0892

\bibitem[\protect\citeauthoryear{{Jaakkola, Salla}, {Sharma, Vivek} \&
  {Annila, Arto}}{{Jaakkola, Salla} et~al.}{2009}]{JaakkolaSalla2009}
{Jaakkola, Salla} {Sharma, Vivek}   {Annila, Arto} 2009, eprint
  arXiv:0906.0254

\bibitem[\protect\citeauthoryear{Kaila \& Annila}{Kaila \&
  Annila}{2008}]{Kaila2008}
Kaila V.~R.,  Annila A.,  2008, Proceedings of the Royal Society A:
  Mathematical, Physical and Engineering Sciences, 464, 3055

\bibitem[\protect\citeauthoryear{{Kardashev, N. S.}}{{Kardashev, N. S.}}{1%
964}]{KardashevN.S.1964}
{Kardashev, N. S.} 1964, Soviet Astronomy, 8

\bibitem[\protect\citeauthoryear{Lal}{Lal}{2008}]{Lal2008}
Lal A.~K.,  2008, Astrophysics and Space Science, 317, 267

\bibitem[\protect\citeauthoryear{{Learned, J. G.}, {Pakvasa, S.},
  {Simmons, W. A.} \& {Tata, X.}}{{Learned, J. G.}
  et~al.}{1994}]{LearnedJ.G.1994}
{Learned, J. G.} {Pakvasa, S.} {Simmons, W. A.}   {Tata, X.} 1994, R.A.S.
  QUARTERLY JOURNAL V. 35

\bibitem[\protect\citeauthoryear{Lineweaver, Fenner \& Gibson}{Lineweaver
  et~al.}{2004}]{GHZ}
Lineweaver C.~H.,  Fenner Y.,    Gibson B.~K.,  2004, Science (New York, N.Y.),
  303, 59

\bibitem[\protect\citeauthoryear{Loeb \& Zaldarriaga}{Loeb \&
  Zaldarriaga}{2007}]{Loeb2007}
Loeb A.,  Zaldarriaga M.,  2007, Journal of Cosmology and Astroparticle
  Physics, 2007, 020

\bibitem[\protect\citeauthoryear{Manning}{Manning}{2006}]{Manning2006}
Manning C.~E.,  2006, American Journal of Science, 306, 303

\bibitem[\protect\citeauthoryear{Miller \& Scalo}{Miller \&
  Scalo}{1979}]{Miller1979}
Miller G.~E.,  Scalo J.~M.,  1979, The Astrophysical Journal Supplement Series,
  41, 513

\bibitem[\protect\citeauthoryear{Mojzsis, Arrhenius, McKeegan, Harrison, Nutman
  \& Friend}{Mojzsis et~al.}{1996}]{Mojzsis1996}
Mojzsis S.~J.,  Arrhenius G.,  McKeegan K.~D.,  Harrison T.~M.,  Nutman A.~P.,
    Friend C.~R.,  1996, Nature, 384, 55

\bibitem[\protect\citeauthoryear{Morris}{Morris}{2006}]{Morris2006R826}
Morris S.~C.,  2006, Current Biology, 16, R826

\bibitem[\protect\citeauthoryear{Nam \& Bozhilov}{Nam \&
  Bozhilov}{2009}]{NamBozhilov2009}
Nam K.,  Bozhilov V.,  2009, in Proceedings, Fifth International Conference
  �Global Changes: Vulnerability, Mitigation And Adaptation� {Intelligence
  and Evolutionary Mechanisms: Origin and Influence on the Ecosystems}.
pp 156--159

\bibitem[\protect\citeauthoryear{Rocha-Pinto, Maciel, Scalo \&
  Flynn}{Rocha-Pinto et~al.}{2000a}]{Rocha_Pinto_SFH}
Rocha-Pinto H.~J.,  Maciel W.~J.,  Scalo J.,    Flynn C.,  2000a, Astronomy and
  Astrophysics, 358, 850

\bibitem[\protect\citeauthoryear{Rocha-Pinto, Maciel, Scalo \&
  Flynn}{Rocha-Pinto et~al.}{2000b}]{Rocha_Pinto_AMR}
Rocha-Pinto H.~J.,  Maciel W.~J.,  Scalo J.,    Flynn C.,  2000b, Astronomy and
  Astrophysics, 358, 869

\bibitem[\protect\citeauthoryear{Sharma \& Annila}{Sharma \&
  Annila}{2007}]{Sharma2007123}
Sharma V.,  Annila A.,  2007, Biophysical Chemistry, 127, 123

\bibitem[\protect\citeauthoryear{{Silagadze, Z. K.}}{{Silagadze, Z. K.}}{2%
008}]{SilagadzeZ.K.2008}
{Silagadze, Z. K.} 2008, Acta Physica Polonica B, 39

\bibitem[\protect\citeauthoryear{Spiegel, Menou \& Scharf}{Spiegel
  et~al.}{2008}]{Spiegel_et_al_08}
Spiegel D.~S.,  Menou K.,    Scharf C.~A.,  2008, The Astrophysical Journal,
  681, 1609

\bibitem[\protect\citeauthoryear{Vukotic \& Cirkovic}{Vukotic \&
  Cirkovic}{2007}]{Vukotic_and_Cirkovic_07}
Vukotic B.,  Cirkovic M.,  2007, Serbian Astronomical Journal, 175, 45

\bibitem[\protect\citeauthoryear{Ward \& Brownlee}{Ward \&
  Brownlee}{2000}]{rare_Earth}
Ward P.,  Brownlee D.,  2000, {Rare Earth : Why Complex Life is Uncommon in the
  Universe}

\bibitem[\protect\citeauthoryear{W\"{u}rtz \& Annila}{W\"{u}rtz \&
  Annila}{2008}]{WurtzPeter2009}
W\"{u}rtz P.,  Annila A.,  2008, Journal of biophysics, 2008, 654

\end{thebibliography}

\end{document}